\documentstyle[12pt]{article}

\begin{document}

\begin{titlepage}

\begin{flushright}
IJS-TP-95/13\\
NUHEP-TH-95-12\\
September 1995\\
\end{flushright}

\vspace{.5cm}

\begin{center}
{\Large \bf $c\to u \gamma$ in Cabibbo suppressed
D meson radiative weak decays\\}

\vspace{1.5cm}

{\large \bf B. Bajc $^{a}$, S. Fajfer $^{a}$ and
Robert J. Oakes $^{b}$\\}

\vspace{.5cm}

{\it a) J. Stefan Institute, University of Ljubljana,
61111 Ljubljana, Slovenia\\}

\vspace{.5cm}

{\it b) Department of Physics and Astronomy,
Northwestern University, Evanston, Il 60208
U.S.A.}

\vspace{1cm}

\end{center}

\centerline{\large \bf ABSTRACT}

\vspace{0.5cm}

We investigate
Cabibbo suppressed $D^{0}$, $D^{+}$  and $D^{+}_{s}$
radiative weak decays in order to find the best mode to test
$c\to u \gamma$ decay. Combining heavy quark effective theory
and the chiral Lagrangian approach we determine  the decay widths.
We calculate
$\Gamma(D^{0}\to \rho^{0}/\omega\gamma)/
\Gamma(D^{0}\to {\bar K}^{*0} \gamma)$ previously  proposed
to search for possible New Physics. However,
we notice that there are large, unknown, corrections
within the Standard Model. We propose a better
alternative, the ratio
$\Gamma(D_{s}^{+}\to K^{*+} \gamma)/
\Gamma(D_{s}^{+}\to \rho^{+} \gamma)$, and
show that it is less sensitive to the
Standard Model.

\end{titlepage}

\setlength {\baselineskip}{0.75truecm}

\noindent
{\bf 1 Introduction}

\vspace{.5cm}

According to the Standard Model, the physics of charm mesons is not as
exciting as the physics of bottom mesons \cite{BIGI1,BIGI2,BIGI3}:
the relevant CKM
matrix elements $V_{cs}$ and $V_{cd}$ are well known,
the $D^{0} -{\bar D}^{0}$ oscillations and CP asymmetries are small,
weak decays of D mesons are difficult to  investigate due
to the strong final state interactions,
and very small branching ratios are expected for rare decays.
However,  authors \cite{BIGI1,BIGI2,BIGI3} have noticed that the
$D^{0} -{\bar D}^{0}$ oscillation and $c \to u \gamma$ decays
obtain contributions coming from non-minimal
supersymmetry which are not present within the Standard Model.
Therefore, these two phenomena might be guides
for a signal of New Physics. In ref. \cite{BIGI2}
it was observed that New Physics can generate $c\to u \gamma$
transitions leading to a deviation from

\begin{equation}
\label{r1}
R_{\rho/\omega}\equiv{\Gamma(D^0 \to \rho^0 /\omega \gamma)
\over\Gamma(D^0 \to {\bar K}^{*0} \gamma)}=
{tan^{2} \theta_{c}\over 2}
\end{equation}

\noindent
(the factor $\frac{1}{2}$
was overlooked in refs. \cite{BIGI1,BIGI2}).
Motivated by the importance of this signal we investigate
Cabibbo suppressed radiative weak decays in which
$c \to u \gamma$ transition occurs.
As a theoretical framework we use a hybrid theory:
a combination of heavy quark effective theory (HQET)
and chiral Lagrangians (CHL) \cite{MW,BD,YCC,G2,BFO}.
This approach, accompanied by the factorization hypothesis,
enables us to use the Standard Model results for
electroweak processes. It is possible to apply other
approaches like for example \cite{burdman}, but the result
which indicates the deviation from $tan^{2} \theta_{c}$
cannot be very different from ours obtained with HQET + CHL.
In fact, our results agree with [9] within the uncertainties.

We calculate the ratios between various
Cabibbo suppressed and Cabibbo allowed
charm meson radiative weak decays.
Analysing them we notice that the
relation (\ref{r1}) can be badly violated
already in the Standard Model framework,
while  a similar relation for $D_{s}^{+}$ radiative
decays, i.e.

\begin{equation}
\label{r2}
R_{K}\equiv{\Gamma(D_s^+ \to K^{*+} \gamma)
\over\Gamma(D_s^+ \to \rho^+ \gamma)}=
tan^{2} \theta_{c}
\end{equation}

\noindent
offers a much better test for $c \to u \gamma$.

The paper is organised as follows: in Sect. 2 we
sketch the relevant theoretical framework for
radiative decays; in Sect. 3 we give results for the
branching ratios of the widths for
Cabibbo suppressed and Cabibbo
allowed radiative decays; finally we make
a short discussion of our results in Sect. 4.

\vspace{.5cm}

\noindent
{\bf 2 Theoretical framework}

\vspace{.5cm}

Experimentally radiative decays of $D$ mesons
have not yet been measured, while the known
branching ratios of $D^{*}$ radiative decays
\cite{pdg,expr} can be described using the
combination HQET + CHL \cite{BFO,CG,ABJ}.

The initial HQET ideas \cite{IW,HG} were implemented
with the chiral Lagrangian formalism for light
pseudoscalar mesons first in \cite{MW,BD,YCC},
and the electromagnetic interaction included in
\cite{CG,ABJ,LLY}. Consequently, the light vector
mesons were introduced \cite{G2}, following the hidden
symmetry approach \cite{BKY}. We will follow the
model described in \cite{BFO}, where in addition to
\cite{G2} the electromagnetic (EM) interaction was introduced.

Let us briefly describe the relevant terms (for the
charm meson radiative weak decays) of the Lagrangian \cite{BFO}.
The main contribution comes from the odd-parity
Lagrangian

\begin{eqnarray}
\label{odd}
{\cal L}_{odd} = &-&4 e {\sqrt 2}
\frac{C_{V\pi\gamma}}{f} \epsilon
^{\mu \nu \alpha \beta}
Tr (\{ \partial_{\mu}{\rho}_{\nu},
\Pi \} Q \partial_{\alpha} B_{\beta})\nonumber\\
&-&4 \frac{C_{VV\Pi}}{f} \epsilon
^{\mu \nu \alpha \beta}Tr (\partial_{\mu}
{\rho}_{\nu} \partial_{\alpha}{\rho}_{\beta} \Pi)\nonumber\\
&-& \lambda^{\prime} e Tr [H_{a}\sigma_{\mu \nu}
F^{\mu \nu} (B) {\bar H_{a}}]\nonumber\\
&+& i {\lambda} Tr [H_{a}\sigma_{\mu \nu}
F^{\mu \nu} (\hat \rho)_{ab} {\bar H_{b}}]\;,
\end{eqnarray}

\noindent
where $C_{VV\Pi} = 0.423$, $C_{V\Pi\gamma} = -3.26\times
10^{-2}$\cite{BOS,FSO}, $f=132$ MeV is the pseudoscalar decay
constant, while the phenomenological parameters
$\lambda$ and $\lambda'$ are constrained
by the analysis \cite{BFO}:

\begin{eqnarray}
\label{lambda}
|\lambda'+{2\over 3}\lambda | & = &(0.50 \pm 0.15)
\;\hbox{GeV}^{-1}\;,\\
|\lambda'-{\lambda\over 3}| & < & 0.19
\;\hbox{GeV}^{-1}\;.
\end{eqnarray}

In (\ref{odd}) $\Pi$ and $\rho_{\mu}$ are the usual pseudoscalar
and vector Hermitian matrices

\begin{eqnarray}
\label{pseudoscalar}
\Pi = \pmatrix{
{\pi^0\over\sqrt{2}}+{\eta_8\over\sqrt{6}}+
{\eta_0\over\sqrt{3}}&\pi^+&K^+\cr
\pi^-&-{\pi^0\over\sqrt{2}}+{\eta_8\over\sqrt{6}}+
{\eta_0\over\sqrt{3}}&K^0\cr
K^-&{\bar K^0}&-2{\eta_8\over\sqrt{6}}+
{\eta_0\over\sqrt{3}}\cr}\;,
\end{eqnarray}

\begin{eqnarray}
\label{vector}
\rho_\mu = \pmatrix{
{\rho^0_\mu + \omega_\mu \over \sqrt{2}} & \rho^+_\mu & K^{*+}_\mu \cr
\rho^-_\mu & {-\rho^0_\mu + \omega_\mu \over \sqrt{2}} & K^{*0}_\mu \cr
K^{*-}_\mu & {\bar K^{*0}}_\mu & \Phi_\mu \cr}
\end{eqnarray}

\noindent
with $\eta=\eta_8\cos{\theta}-\eta_0\sin{\theta}$,
$\eta'=\eta_8\sin{\theta}+\eta_0\cos{\theta}$ and
$\theta=-23^o$ \cite{pdg} is the
$\eta-\eta'$ mixing angle.
$Q=diag(2/3,-1/3,-1/3)$ is the light quark sector
charge matrix,

\begin{equation}
\label{defha}
H_a={1 \over 2} (1 + \!\!\not{\! v}) (\sqrt{m_{D^{a*}}}
D_\mu^{a*}\gamma^{\mu} - \sqrt{m_{D^a}} D^a \gamma_{5})\;,
\end{equation}

\noindent
where $D_\mu^{a*}$ and $D^a$ annihilate,
respectively, a spin-one and spin-zero meson
$c{\bar q}^{a}$ ($q^a=u$, $d$ or $s$)
of velocity $v_{\mu}$
and ${\bar H}_a\equiv\gamma^0H_a^\dagger\gamma^0$.
Finally, $F_{\mu\nu}({\hat \rho})=
\partial_\mu{\hat\rho}_\nu-
\partial_\nu{\hat\rho}_\mu+
[{\hat\rho}_\mu,{\hat\rho}_\nu]$,
${\hat\rho}_\mu=ig_V\rho_\mu/\sqrt{2}$ with
$g_V=5.8$ \cite{G2}, and
$F_{\mu\nu}(B)=\partial_\mu B_\nu-
\partial_\nu B_\mu$ with $B_\mu$ being the
photon field with the EM coupling constant $e$.

The first (third) term in (\ref{odd}) describes
the anomalous direct emission of the photon by the
light (heavy) meson, while the second (fourth) term,
together with the vector meson dominance (VMD)
coupling

\begin{eqnarray}
\label{vmd}
{\cal L}_{V-\gamma} & = & - m_V^2 {e \over g_V} B_{\mu}
(\rho^{0\mu} + {1 \over 3} \omega^{\mu} -
{\sqrt{2} \over 3} \Phi^{\mu})
\end{eqnarray}

\noindent
describes a two step photon emission, with an
intermediate neutral vector meson with mass
$m_V$ which transforms to the final photon.

A charged charm meson can emit a real photon
also through the usual electromagnetic coupling

\begin{equation}
\label{emheavy}
{\cal L}_{EM}=-e v^\mu B_\mu Tr[H_a(Q-2/3)_{ab}{\bar H}_b]\;,
\end{equation}

\noindent
while a charged light vector meson can produce through

\begin{equation}
\label{vvv}
{\cal L}_{VVV}={1\over 2 g_V^2}tr[
F_{\mu\nu}({\hat\rho})
F^{\mu\nu}({\hat\rho})]
\end{equation}

\noindent
first a neutral vector meson, which subsequently
transforms via VMD (\ref{vmd}) to a photon.

The weak Lagrangian is approximated by the current-current
type interaction

\begin{equation}
\label{fermi}
{\cal L}_{W}^{eff}(\Delta c = 1) =
-{G_F\over\sqrt{2}}
[a_{1}({\bar u}d')^\mu({\bar s}'c)_\mu+
a_{2}({\bar s}'d')^\mu({\bar u}c )_\mu]\;,
\end{equation}

\noindent
where $({\bar q}_1 q_2)^\mu\equiv
{\bar q}_1\gamma^\mu(1-\gamma^5)q_2$, $G_F$ is
the Fermi constant and
$a_{1,2}$ are the QCD Wilson coefficients,
which depend on the energy scale $\mu$.
One expects the scale to be the heavy quark mass
and we take $\mu \simeq 1.5$ GeV which
gives $a_{1} = 1.2$ and  $a_{2} = -0.5$,
with an approximate $20\%$ error. Since we are
interested only in the physics of the first two generations,
we can express the weak eigenstates $d'$, $s'$ with
the mass eigenstates $d$, $s$ using the Cabibbo angle
instead of the CKM matrix:

\begin{eqnarray}
\label{dstransf}
\pmatrix{ d' \cr
          s' \cr } =
\pmatrix{ \cos{\theta_c} & -\sin{\theta_c} \cr
          \sin{\theta_c} &  \cos{\theta_c} \cr }
\pmatrix{ d \cr
          s \cr }
\end{eqnarray}

\noindent
with $\sin{\theta_c}=0.222$. Possible contributions caused
by the penguin type diagrams are found to
be very small \cite{burdman}.

Many heavy meson weak nonleptonic
amplitudes \cite{KXC,WSB1,G4} have been
calculated using the factorization approximation.
In this approach the quark currents are
approximated by the ``bosonised" currents
\cite{MW,G2,BFO}, of which only

\begin{eqnarray}
\label{ethirty}
({\bar q}^a c)_\mu&=&i(m_{D^{*a}}f_{D^{*a}}D^{*a}_\mu-
m_{D^a}f_{D^a}v_\mu D^a)\;,\\
({\bar q}_b q_a)^{\mu}&=&-f\partial^\mu\Pi_{ab}+
m_V f_V\rho^\mu_{ab}
\end{eqnarray}

\noindent
will contribute to our amplitudes. The
numerical values for the masses will be taken
from \cite{pdg} and for the decay constants from
\cite{G4}.

It is now straightforward to calculate the decay widths.
The result, of course, depends on the
numerical values we take for $(\lambda' +
2 \lambda /3)$ and $(\lambda'-\lambda/3)$.

\vspace{.5cm}

\noindent
{\bf 3 Cabibbo suppressed radiative weak decays in HQET + CHL}

\vspace{.5cm}

Apart from the Cabibbo allowed decays
$D^0\to{\bar K}^{*0}\gamma$ and $D_s^+\to\rho^+\gamma$,
five once Cabibbo suppressed ($D^{0}\to\rho^{0}\gamma$,
$D^{0}\to\omega\gamma$, $D^{0}\to\phi\gamma$,
$D^{+}\to\rho^{+} \gamma$, $D_s^{+}\to K^{*+} \gamma$)
and two doubly Cabibbo suppressed ($D^0\to K^{*0}\gamma$
and $D^+\to K^{*+}\gamma$) decays are possible.

According to \cite{WSB1}
the weak amplitudes can be categorized into
two groups: quark decays and weak annihilations.
As these authors have noticed, the factorization
works much better for the quark decays.
The decays of $D_s^+$ and $D^+$ are of the
quark decay type (their amplitudes are proportional
to $a_1$, see below), and therefore the results for
these decays are more trustworthy than for the $D^0$
$D$ decays, which proceed through weak annihilation
diagrams (proportional to $a_2$).

We write the amplitude for the $D^{q} \to V^{q} \gamma$,
where $q$ stands for the charge of D meson
($q= 1$ stands for $+$ charge, while $q = 0$ is for
the neutral D mesons)

\begin{eqnarray}
\label{ampl}
A(D^{q} &\to &V^{q} \gamma)  =
e{G_F\over\sqrt 2} K_c a(q)
[ C^{(1)}_{D V \gamma} \epsilon _{\mu \nu \alpha \beta}
k^{\mu} \epsilon_{\gamma}^{\nu *} v^{\alpha} \epsilon_{V}^{\beta *}
\nonumber\\
& + &iC^{(2)}_{D V \gamma} m_{V} ( \epsilon_{\gamma}^{*}
\cdot \epsilon_{V}^{*}  - \frac{\epsilon_{\gamma}^{*} \cdot p_{V}
\epsilon_{V}^{*}\cdot k}
{p_{V} \cdot k} )]
\end{eqnarray}

\noindent
with $a(+1)=a_{1}$ and $a(0)=a_{2}$. $(k,\epsilon_\gamma)$
and $(p_V,\epsilon_V)$ are the $4$-momenta and polarization
vectors of the photon and vector meson respectively, while
$v$ is the $4$-velocity of the heavy meson.

The overall factor
$K_c$ contains the Cabibbo angle and is equal to
$\cos{\theta_c}^2$ for allowed decays, to
$+\sin{\theta_c}\cos{\theta_c}$ (when there is
no $s$ quark or antiquark in the final $V$)
or $-\sin{\theta_c}\cos{\theta_c}$ (when there is
at least one $s$ quark or antiquark in the final $V$)
for once suppressed
decays and to $-\sin{\theta_c}^2$ for double
suppressed decays. The coefficients $C^{(i)}$ in (\ref{ampl})
can be written as

\begin{eqnarray}
\label{c1}
C^{(1)}_{D V \gamma} & = &
( C_{VV\Pi} \frac{1}{g_{V}} + C_{V \Pi \gamma})
4 {\sqrt 2} f_{D} m_{D}^{3} b^V\nonumber\\
&+ &4 [\lambda^{\prime} + (\frac{2}{3} -q) \lambda] f_{D^{*}} f_{V}
\frac{m_{D^{*}}m_{V}}{m_{D^{*}}^{2} - m_{V}^{2}}
\sqrt{m_{D}m_{D^{*}}}b_0^V\;,\\
C^{(2)}_{D V \gamma} & = & q f_D f_V\;.
\end{eqnarray}

The coefficient $b^V$ is equal to $(2/3-q)/(m_D^2-m_P^2)$
for $V=({\bar K}^{*0}$, $K^{*0}$, $\rho^+$, $K^{*+})$,
for which $P=({\bar K}^0$, $K^0$, $\pi^+$, $K^+)$.
For the remaining final state vector mesons this
coefficient is expressed as

\begin{equation}
\label{defbv}
b^V=\sum_{i=1}^{3} \frac{b^{VP_i}}{m_{D}^{2} - m_{P_{i}}^{2}}\;,
\end{equation}

\noindent
where the pole coefficients $b^{VP_i}$ are given in Table 1.
Furthermore we have $b_0^V=-1/\sqrt{2}$ for $V=\rho^0$,
$b_0^V=1/\sqrt{2}$ for $V=\omega$ and $b_0^V=1$ otherwise.

In ref. \cite{BIGI1,BIGI2} it was noticed that
a nice bonus can be obtained by
measuring the charm meson decay width
$D \to \rho / \omega \gamma$ which
is generated by $c\to u \gamma$ transitions.
Namely, the authors claim that observing the
violation of equation (\ref{r1}) would then
eventually signal New Physics \cite{BIGI1}.
Using our model, which pretends to describe the low energy
meson physics within the Standard Model, we find that
this relation does not exactly hold
due to U(3) breaking. We assume that
the leading effect of this breaking is to change
the values of the masses and decay constants
for different members of the same multiplets
and between octet and singlet. However, one
would naively expect deviations from this limit
in the Standard Model of the order of
$20-30\%$. We will see that this is not true for
the $D^0\to\rho^0/\omega\gamma$ decay, but it is
correct for $D_s^+$ Cabibbo suppressed radiative
weak decay.

Within our framework $\lambda$ and
$\lambda^{\prime}$ are the most important parameters for
charm meson radiative decays \cite{BFO},
and therefore we present the ratios
of the decay widths as functions of combinations
of $\lambda$ and $\lambda^{\prime}$.
Our result for $R_\rho$ (\ref{r1})
is showed on Fig. 1: if the combination of
$(\lambda^{\prime} + \frac{2}{3} \lambda)$
turns out to be negative, the ratio $R_\rho$
can approach $0$. As it is known from
$D^{0} \to {\bar K}^{*0} \gamma$ \cite{BFO},
the negative values $(\lambda^{\prime} + \frac{2}{3} \lambda)$
cause a destructive interference between the photon emission
by the heavy meson and the photon emission by the light meson.
A similar effect is possible also in the decay
$D^{0} \to \rho^0 \gamma$, only that
the $0$ is achieved at a different value of
$(\lambda^{\prime} + \frac{2}{3} \lambda)$, because
the model parameters are here slightly different
due to $U(3)$ breaking. It is obvious that such
a large sensitivity to the model parameters
does not allow us to conclude anything about some
New Physics. If $(\lambda^{\prime} + \frac{2}{3} \lambda)$
turns out to be positive, the decays are much
easier to detect experimentally, and also the theoretical
situation is clearer, since the curve is approaching
the ideal theoretical value. A large disagreement
with the theoretical prediction (\ref{r1})
would give in this case some sign of New Physics.
But even here one should be careful, since
in this case the amplitudes are
approximately proportional to the decay constants
of the final vector meson.
This can be seen, if we calculate the decay $D^{0} \to \omega \gamma$
with the values of the light vector decay constants
taken from \cite{G4} :
$f_{K*} = f_{\rho} = 221$ MeV and $f_{\omega} = 156$ MeV.
In this case we get for $R_\omega$ a
similar curve as in Fig. 1, but for large positive
values $(\lambda^{\prime} + \frac{2}{3} \lambda)$ the
ratio is approaching a value of
approximately $0.5$ instead of 1.
The fact can be explained by the difference in
the decay constants, i.e. $(f_{\omega}/f_{K*})^{2} \simeq 0.5$.

The ratio $\Gamma(D^0\to\Phi\gamma)/\Gamma(D^0\to
{\bar K}^{*0}\gamma)$ would indicate the deviation from
$\tan{\theta_c}^2$ instead of $\tan{\theta_c}^2/2$
like for the $\rho,\omega$ case. When calculated,
it exhibit a similar behaviour like $D^0\to\omega\gamma$,
and therefore we find it is not useful to understand
$c\to u\gamma$ physics.

The decays $D^+\to\rho^+\gamma$
is also not of great importance in our purpose to find
New Physics, since the $D^+$ does not have
Cabibbo allowed decays.

Contrary to the above cases we find that the decay $D_{s}^{+} \to
K^{*+} \gamma$ offers a much better chance to test New Physics.
Using the general formulas for the amplitudes (\ref{ampl}) it
is easy to derive a deviation from equation (\ref{r2}),
which is exactly correct only in the U(3) limit.
The result for $R_K$ as a function of
$(\lambda^{\prime} -\frac{1}{3} \lambda)$ is presented
on Fig. 2 (note the changed scale with respect to
Fig. 1). We notice that the result is rather stable within
the allowed region for $(\lambda^{\prime} -\frac{1}{3} \lambda)$.
The discrepancy to relation (\ref{r2}) is due to U(3) breaking
and is of order $30\%$, as usually expected. If the
the experimental
results are found to be far away from the curve  Fig. 2,
one can interpret it as a sign of New Physics.

We point out that it is difficult to observe
all these decays. In fact the Cabibbo allowed
decays are already rare: the branching ratio for
$D^0\to{\bar K}^{*0}\gamma$ is smaller than
$0.3\times 10^{-4}$ for $(\lambda'+2\lambda /3)<0$
and around $(2-4)\times10^{-4}$
for $(\lambda'+2\lambda /3)>0$, while for
$D_s^+\to\rho^+\gamma$ the branching ratio
is around $(2-7)\times 10^{-4}$ \cite{BFO}.

\vspace{0.5cm}

\noindent
{\bf 4 Conclusions}

\vspace{0.5cm}

We determine amplitudes of Cabibbo suppressed radiative decays
using the combination of
heavy quark symmetry and chiral symmetry, which builds an effective
strong, weak and electromagnetic Lagrangian.
This theoretical framework just illustrates the characterictics
of these amplitudes in the Standard Model.
In our framework two parameters, $\lambda$ and
$\lambda^{\prime}$, are not well known.
We show the dependence of the ratio between the Cabibbo
suppressed and Cabibbo allowed decay widths
on the combination of
$\lambda$ and $\lambda^{\prime}$.  We find that
is better to search for a signal of New Physics
coming from $c \to u \gamma$ decays from the ratio
$\Gamma (D^{+}_{s} \to K^{*+} \gamma)/
\Gamma (D^{+}_{s} \to \rho^{+} \gamma)$
instead of the proposed ratio
$\Gamma(D^{0} \to \rho^{0}/\omega \gamma) /
\Gamma(D^{0} \to {\bar K}^{*0} \gamma) $
\cite{BIGI1,BIGI2,BIGI3}.

\vskip 0.5cm
{\it Acknowledgement.} This work was supported in part by the
Ministry of Science and Technology of the Republic
of Slovenia (B.B. and S.F.)
and by the U.S. Department
of Energy, Division of High Energy Physics,
under grant No. DE-FG02-91-ER4086 (R.J.O.).

\newpage

\centerline{FIGURES}

\noindent
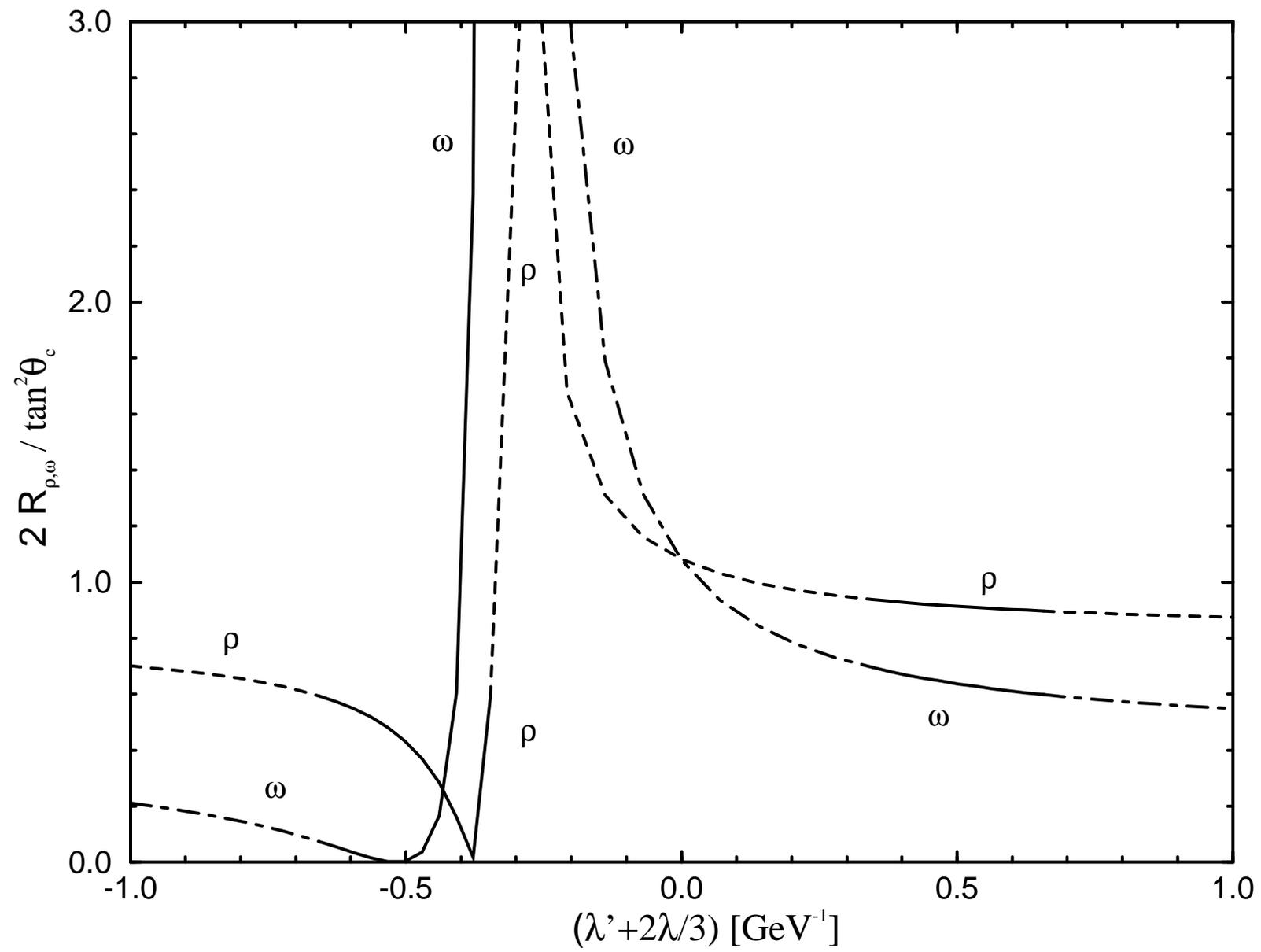
Fig. 1: The ratio $2 R_{\rho/\omega}/\tan{\theta_c}^2$
as function of the combination $\lambda'+2\lambda/3$. The full
(dashed/dot-dashed) lines denote the experimentally
allowed (forbidden) values for this combination.
In the U(3) symmetry limit
of the Standard Model this ratio is equal $1$.

\vskip 0.5cm
\noindent
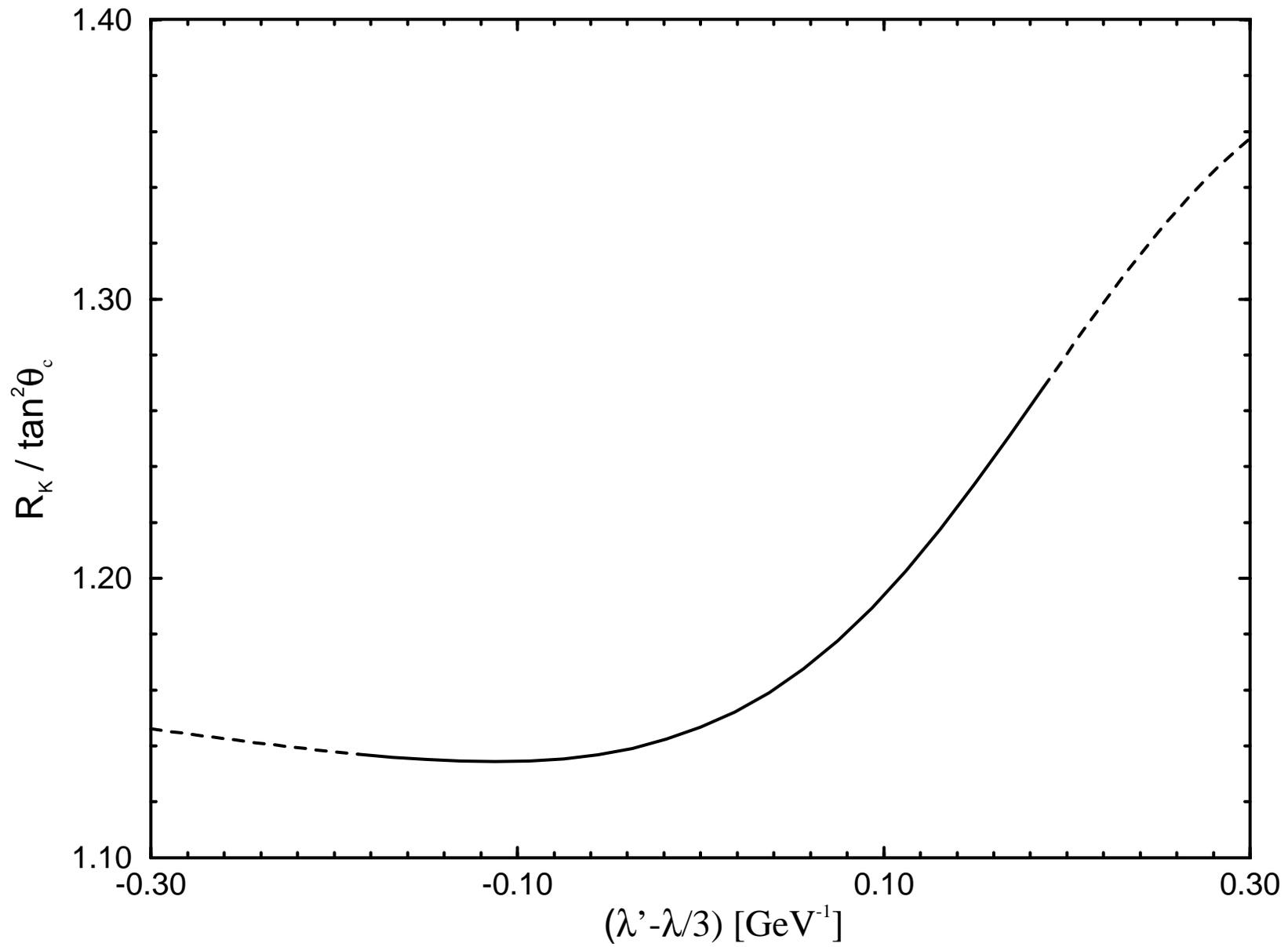
Fig. 2: The ratio $R_K/\tan{\theta_c}^2$
as function of the combination $\lambda'-\lambda/3$. The full
(dashed) lines denote the experimentally allowed (forbidden)
values for this combination. In the U(3) symmetry limit
of the Standard Model this ratio is equal $1$.

\newpage

\begin{table}[h]
\begin{center}
\begin{tabular}{|c||c|c|c|}
\hline
$\enspace$ & $\pi^{0}$ & $ \eta$ & $\eta^{\prime}$ \\
\hline
\hline
$\rho^0$
& $\frac{1}{3\sqrt{2}}$
& $ -\frac{1}{\sqrt{2}}c(c-\sqrt{2} s)$
& $ -\frac{1}{\sqrt{2}}s(\sqrt{2}c+s)$\\
\hline
$\omega$
& $\frac{1}{{\sqrt 2}}$
& $ -\frac{1}{3\sqrt{2}}c(c-\sqrt{2}s)$
& $ -\frac{1}{3\sqrt{2}}s(\sqrt{2}c+s)$\\
\hline
$\phi$
& $0$
& $ \frac{\sqrt{2}}{3}c(\sqrt{2}c+s)$
& $ -\frac{\sqrt{2}}{3}s(c-\sqrt{2}s)$\\
\hline
\end{tabular}
\end{center}
\caption{The $b^{VP_i}$ coefficents defined in relation
(19), where $s=\sin{\theta}$, $c=\cos{\theta}$ and
$\theta$ is the $\eta-\eta'$ mixing angle.}
\end{table}


\newpage

\begin{thebibliography}{99}
\bibitem{BIGI1} I. Bigi, F. Gabbiani,
A. Masiero, Z. Phys. {\bf C48} (1990) 633.
\bibitem{BIGI2} I.I. Bigi, Report No. CERN-TH.7370/94,
Report No. UND-HEP-94-BIG08, hep-ph/9408235 (unpublished).
\bibitem{BIGI3} I.I. Bigi, Report No. UND-HEP-95-BIG08,
hep-ph/9508294 (unpublished).
\bibitem{MW} M. Wise, Phys. Rev. {\bf D45} (1992) 2188.
\bibitem{BD} G. Burdman and J. Donoghue,
Phys. Lett.{\bf B280} (1992) 287.
\bibitem{YCC} T.M. Yan, H.Y. Cheng,
C.Y. Cheung, G.L. Lin, Y.C. Lin, H.L. Yu,
Phys. Rev. {\bf D46} (1992) 1148.
\bibitem{G2} R. Casalbuoni, A. Deandra, N.Di Bartolomeo, R. Gatto,
F. Feruglio, G. Nardulli, Phys. Lett. {\bf B292} (1992) 371.
\bibitem{BFO} B. Bajc, S. Fajfer and R.J. Oakes,
Phys. Rev {\bf D51} (1995) 2230.
\bibitem{burdman} G. Burdman, E. Golowich, J.L. Hewett and
S. Pakvasa, Report No. FERMILAB-Pub-94/412-T,
UMHEP-415, SLAC-PUB-6692, UH-511-811-94, hep-ph/9502329(unpublished).
\bibitem{pdg} Review of Particle Properties 1994,
Phys. Rev. {\bf D 50} (1994) 1173.
\bibitem{expr} CLEO Collab., F.Butler at al.,
Phys. Rev. Lett. {\bf 69} (1992) 2041.
\bibitem{CG} P. Cho and H. Georgi,
Phys. Lett. {\bf B296} (1992) 408.
\bibitem{ABJ} J. Amundson, C.G. Boyd,
E. Jenkins, M. Luke, A. Manohar,
J. Rosner, M. Savage and M. Wise,
Phys. Lett. {\bf B296} (1992) 415.
\bibitem{IW} N. Isgur and M. Wise, Phys. Lett.{\bf B232}
(1989) 113; {\bf B237} (1990) 527.
\bibitem{HG} H. Georgi, Phys. Lett. {\bf B240} (1990) 447.
\bibitem{LLY}  H.Y. Cheng, C.Y. Cheung,
G.L. Lin, Y.C. Lin, T.M. Yan and H.L. Yu, Phys. Rev.
{\bf D47} (1993) 1030; {\bf D49} (1994) 2490.
\bibitem{BKY} M. Bando, T. Kugo, S. Uehara,
K. Yamawaki and T. Yanagida, Phys. Rev. Lett.
{\bf 54} (1985) 1215;
M. Bando, T. Kugo, and K. Yamawaki,
Nucl. Phys {\bf B259} (1985) 493;
Phys. Rep. 164 (1988) 217.
\bibitem{BOS} E. Braaten, R.J. Oakes and Sze-Man Tse,
Int. Jour. Mod. Phys. {\bf A5} (1990) 2737.
\bibitem{FSO} S. Fajfer, K. Suruliz and R.J. Oakes,
Phys. Rev. {\bf D46}(1992) 1195.
\bibitem{KXC} A.N. Kamal, Q.P. Xu and A. Czarnecki,
Phys. Rev. {\bf D49}(1994) 1330.
\bibitem{WSB1} M. Bauer, B. Stech and M. Wirbel,
Z. Phys. {\bf C34} (1987)103.
\bibitem{G4} A. Deandra, N. Di Bartolomeo,
R. Gatto and G. Nardulli, Phys. Lett. {\bf B318} (1993) 549.
\end{thebibliography}
\end{document}